\documentstyle[aps,preprint]{revtex}

\tightenlines

\begin{document}
\title{Theory of Sinusoidal Modulation of the Resonant Neutron Scattering in High \\
Temperature Superconductors}
\author{Tao Li}
\address{Center for Advanced Study, Tsinghua University, \\
P. R. China,100084}
\maketitle

\begin{abstract}
A model with interlayer pairing is proposed to explain the sinusoidal
modulation of the resonant neutron scattering in high temperature
superconductors. It is found that the interlayer pairing has s-wave symmetry
in the CuO$_2$ plane and has comparable magnitude with the d-wave intralayer
pairing. It is also found that the interlayer pairing mainly affects
momentum close to the hot spots on the Fermi surface while its effect on the
gap nodes is negligible. It is pointed out that these characteristics of the
interlayer pairing can be understood in a model in which the superconducting
pairing originates from the exchange of the antiferromagnetic spin
fluctuation.
\end{abstract}

\newpage

The discovery of the resonant neutron scattering is one of the most
important progress in the high-T$_c$ field in recent years\cite{1,2,3,4,5}.
This resonance, also called $\left( \pi ,\pi \right) $ resonance, has many
interesting characteristics and has attracted much theoretical attention\cite
{6,7,8,9,10,11,12,13,14,15,16,17,18}. The c-axis modulation is an
interesting and intriguing problem in the explanation of the $\left( \pi
,\pi \right) $ resonance. In experiments, only the odd channel magnetic
response $\left( q_c=\pi \right) $ has been observed in the CuO$_2$ bilayer
systems (the sinusoidal modulation), while in most single CuO$_2$
plane-based RPA theories, the interlayer exchange is far too small to
distinguish the even from the odd channel\cite{11,12,13,16,18}.

Physically, the momentum dependence of the magnetic response is closely
related to the internal structure of the superconducting Cooper pair. For
point-like s-wave pairing, the magnetic response is generally momentum
independent and is suppressed due to the singlet nature of the pair. While
for d-wave pairing which takes place between nearest neighboring sites on a
plane, the magnetic response is strongly momentum dependent. In fact, here
we can take the Cooper pair roughly as the coherent superposition of two
antiferromagnetic spin configuration. Thus, if we look at the pair with a
momentum transfer of the order $\left( \pi ,\pi \right) $, we will find an
enhanced magnetic response. Similarly, if there exists interlayer pairing
between the two CuO$_2$ planes in the CuO$_2$ bilayer, the odd channel
magnetic response $\left( q_c=\pi \right) $ will be enhanced while the even
channel response $\left( q_c=0\right) $ will be suppressed. Although such
pairing is obviously favored by the interlayer exchange coupling, it is
totally neglected in the single CuO$_2$ plane-based theories.

In this letter, we find the sinusoidal modulation of the $\left( \pi ,\pi
\right) $ resonance can be naturally explained with the inclusion of the
interlayer pairing. We find that the interlayer pairing has s-wave symmetry
in the CuO$_2$ plane and has comparable magnitude with the intralayer d-wave
pairing. We find that the interlayer pairing mainly affects momentum close
to the hot spots on the Fermi surface and has negligible effect on the gap
nodes.

To model the CuO$_2$ bilayer with interlayer pairing, we take the following
mean field Hamiltonian\cite{19} 
\begin{equation}
H_{MF}=\sum\limits_{k,n,\sigma }\xi _kc_{k\sigma }^{(n)\dagger }c_{k\sigma
}^{(n)}+\sum\limits_{k,n}(\Delta _kc_{k\uparrow }^{(n)\dagger
}c_{-k\downarrow }^{(n)\dagger }+h.c.)+\sum\limits_k\left[ \Delta
_k^{^{\prime }}(c_{k\uparrow }^{(1)\dagger }c_{-k\downarrow }^{(2)\dagger
}+c_{k\uparrow }^{(2)\dagger }c_{-k\downarrow }^{(1)\dagger })+h.c.\right] 
\end{equation}
in which n=1,2 is the layer index. $\xi _k$ is the dispersion in the CuO$_2$
plane. Here we use the dispersion derived from fitting the ARPES result in Bi%
$_2$Sr$_2$CaCu$_2$O$_8$\cite{18,20,21}, $\xi _k=-t(\cos k_x+\cos
k_y)-t^{\prime }\cos k_x\cos k_y$ $-t^{\prime \prime }(\cos 2k_x+\cos
2k_y)-t^{\prime \prime \prime }(\cos 2k_x\cos k_y+\cos k_x\cos
2k_y)-t^{\prime \prime \prime \prime }\cos 2k_x\cos 2k_y-\mu ,$ $t=0.2975$
eV, $t^{\prime }=-0.1636$ eV, $t^{\prime \prime }=0.02595$ eV, $t^{\prime
\prime \prime }=0.05585$ eV, $t^{\prime \prime \prime \prime }=-0.0510$ eV, $%
\mu $ is the chemical potential. Note that we have neglected the interlayer
hopping term in the dispersion since no band splitting is observed in
experiment. $\Delta _k=\frac{\Delta _0}2(\cos k_x-\cos k_y)$ is the
intralayer d-wave pairing function. $\Delta _k^{^{\prime }}$ is the
interlayer pairing function. As will be shown later, $\Delta _k^{^{\prime }}$
has strong momentum dependence in the CuO$_2$ plane. However, such momentum
dependence is not essential for the discussion of the $\left( \pi ,\pi
\right) $ resonance since the low energy magnetic response at $q=\left( \pi
,\pi \right) $ is determined mainly by the electronic transition between the
hot spots on the Fermi surface (see FIG. 1). For the sake of simplicity, we
will take $\Delta _k^{^{\prime }}$ as momentum independent for the moment.
The relative phase between $\Delta _k^{^{\prime }}$ and $\Delta _k$ is
another important issue. In the absence of the interlayer hopping, $\Delta
_k^{^{\prime }}$ can be either real or purely imaginary to meet the
requirement of the time reversal symmetry. However, a real $\Delta
_k^{^{\prime }}$ will lead to different energy gap at momentum ($k_x,k_y$)
and ($k_y,k_x$), since $\Delta _k^{^{\prime }}$ and $\Delta _k$ have
different momentum dependence in the CuO$_2$ plane. Therefore $\Delta
_k^{^{\prime }}$ must be purely imaginary.

To discuss the c-axis modulation of the CuO$_2$ bilayer system, it is
convenient to use the bonding band and the anti-bonding band representation%
\cite{6,11,16} 
\begin{equation}
\begin{array}{l}
c_k^{(b)}=\frac 1{\sqrt{2}}(c_k^{(1)}+c_k^{(2)}) \\ 
c_k^{(a)}=\frac 1{\sqrt{2}}(c_k^{(1)}-c_k^{(2)})
\end{array}
\end{equation}
Here b and a represent the bonding and the anti-bonding band respectively.
In this representation, the mean field Hamiltonian reads, 
\begin{equation}
H_{MF}=\sum\limits_{k,\alpha ,\sigma }\xi _kc_{k\sigma }^{(\alpha )\dagger
}c_{k\sigma }^{(\alpha )}+\sum\limits_{k,\alpha }(\Delta _k^{(\alpha
)}c_{k\uparrow }^{(\alpha )\dagger }c_{-k\downarrow }^{(\alpha )\dagger
}+h.c.)
\end{equation}
$\alpha $=a,b is the band index, $\Delta _k^{(\alpha )}=\Delta _k+f(\alpha
)\Delta _k^{^{\prime }}$, where 
\[
f(\alpha )=\left\{ 
\begin{array}{l}
+1,\text{ for }\alpha =\text{b} \\ 
-1,\text{ for }\alpha =\text{a}
\end{array}
\right. 
\]
In the bonding and the anti-bonding band representation, the even and the
odd channel magnetic response come from the intraband and the interband
electronic transition respectively\cite{6,11,16} 
\begin{equation}
\begin{array}{l}
\chi _0^{(even)}(q,\omega )=\chi _0^{(aa)}(q,\omega )+\chi
_0^{(bb)}(q,\omega ) \\ 
\chi _0^{(odd)}(q,\omega )=\chi _0^{(ab)}(q,\omega )+\chi _0^{(ba)}(q,\omega
)
\end{array}
\end{equation}
in which the mean field susceptibility $\chi _0^{(aa)}(q,\omega ),\chi
_0^{(bb)}(q,\omega ),\chi _0^{(ab)}(q,\omega )$ and $\chi _0^{(ba)}(q,\omega
)$ are given by ( for simplicity we discuss the zero temperature case)\cite
{11,22} 
\begin{equation}
\chi _0^{(ij)}(q,\omega )=\frac 14\sum\limits_k(1-\frac{\xi _k\xi
_{k+q}+\Delta _k^{(i)}\Delta _{k+q}^{(j)*}}{E_k^{(i)}E_{k+q}^{(j)}})(\frac 1{%
\omega +E_k^{(i)}+E_{k+q}^{(j)}+i\delta }-\frac 1{\omega
-E_k^{(i)}-E_{k+q}^{(j)}+i\delta })
\end{equation}
here i, j=a, b, $E_k^{(i)}=\sqrt{\left( \xi _k\right) ^2+\left| \Delta
_k^{(i)}\right| ^2}$ is the quasiparticle energy.

To see the effect of the interlayer pairing on the momentum dependence of
the magnetic response, let us look at the BCS coherence factor, $(1-\frac{%
\xi _k\xi _{k+q}+\Delta _k^{(i)}\Delta _{k+q}^{(j)*}}{E_k^{(i)}E_{k+q}^{(j)}}%
)$, which contains the information about the internal structure of the
Cooper pair. In the absence of the interlayer pairing, the even and the odd
channel have the same coherence factor $(1-\frac{\xi _k\xi _{k+q}+\Delta
_k\Delta _{k+q}}{E_kE_{k+q}})$ and both channels are fully enhanced at
momentum transfer $q=\left( \pi ,\pi \right) $ since $\Delta _k\Delta
_{k+q}<0$. That is, the system dose not distinguish the even and the odd
channel at the mean field level. As we have mentioned at the beginning of
this paper, the RPA correction from the interlayer exchange is far too small
to make this mean field result agree with the observed large difference
between the even and the odd channel.

In the presence of the interlayer pairing, the even and the odd channel
behave differently. For the odd channel, since $\Delta _k^{(a)}\Delta
_{k+q}^{(b)*}=\Delta _k^{(b)}\Delta _{k+q}^{(a)*}=-\left( \Delta _k^2+\left|
\Delta _k^{^{\prime }}\right| ^2\right) $(using the properties $\Delta
_{k+q}=-\Delta _k$, $\Delta _{k+q}^{^{\prime }}=\Delta _k^{^{\prime }}$),
the coherence factor can still reach its maximum value 2 on the Fermi
surface. That is, the odd channel magnetic response is still fully enhanced.
While for the even channel, since $\Delta _k^{(a)}\Delta
_{k+q}^{(a)*}=-\left( \Delta _k^2-\left| \Delta _k^{^{\prime }}\right|
^2-2i\Delta _k\Delta _k^{^{\prime }}\right) =-\left( \Delta _k^2-\left|
\Delta _k^{^{\prime }}\right| ^2\right) =\Delta _k^{(b)}\Delta _{k+q}^{(b)*}$
(the cross term of $\Delta _k$ and $\Delta _k^{^{\prime }}$ vanishes upon
summing over $k$ since $\Delta _k$ and $\Delta _k^{^{\prime }}$ have
different symmetry), the coherence factor can take any value ranging from 0
(totally suppressed) to 2 (fully enhanced) on the Fermi surface depending on
the ratio$\frac{\left| \Delta _k^{^{\prime }}\right| }{\Delta _0}$. As a
result, the even channel magnetic response is suppressed with the increase
of the interlayer pairing. FIG. 2. shows the calculated susceptibility for $%
\frac{\left| \Delta _k^{^{\prime }}\right| }{\Delta _0}=1$(the magnitude of
the interlayer pairing will be discussed later). We see the interlayer
pairing suppresses the even channel magnetic response very effectively.

So far, we have discussed the bare susceptibility $\chi _0$. To obtain the
fully renormalized susceptibility, we still have to include the RPA
correction from the antiferromagnetic exchange coupling. In the presence of
the interlayer exchange coupling, the even and odd channel magnetic
responsee are renormalized differently\cite{10,11,16}, 
\begin{equation}
\begin{array}{l}
\chi ^{(even)}(q,\omega )=\frac{\chi _0^{(even)}(q,\omega )}{1+(J_q+J_p)\chi
_0^{(even)}(q,\omega )/2} \\ 
\chi ^{(odd)}(q,\omega )=\frac{\chi _0^{(odd)}(q,\omega )}{1+(J_q-J_p)\chi
_0^{(odd)}(q,\omega )/2}
\end{array}
\end{equation}
in which $J_q=J(\cos q_x+\cos q_y)$ is the intralayer exchange, and $J_p$ is
the interlayer exchange coupling. Experimentally, $J\sim $0.15 eV, $%
J_p/J\sim $ 0.1\cite{23}. FIG. 3 shows the renormalized susceptibility.
After the RPA correction, a sharp resonance appears well below the gap edge
in the odd channel\cite{18}. While in the even channel, there is only a
small peak very close to the gap edge. This result can be understood by
examining the resonance condition for both channels. As can be seen from
FIG. 2, the resonance condition is fulfilled well below the gap edge in the
odd channel. While in the even channel, this condition is only fulfilled
very close to the gap edge because of the reduced magnitude of the bare
susceptibility in the even channel( $J_p$ alone is too small to produce the
observed even-odd difference). Here we find the magnetic response starts at
different energies in the odd and even channel. This agrees very well with
experimental observations\cite{24}. According to our theory, the resonance
energy of the odd channel is unrelated to the superconducting gap while the
energy threshold for even channel magnetic response is about twice of the
superconducting gap.

In the foregoing discussion, we have neglected the momentum dependence of
the interlayer pairing. This is reasonable for the discussion of the $\left(
\pi ,\pi \right) $ resonance which mainly concerns the hot spots on the
Fermi surface. However, a momentum independent interlayer pairing is
inconsistent with the experimental observation of the gap nodes along the $%
\left( 0,0\right) -\left( \pi ,\pi \right) $ direction since the total
energy gap equals $\sqrt{\Delta _k^2+\left| \Delta _k^{^{\prime }}\right| ^2}
$. To be consistent with the existence of the gap nodes, $\Delta
_k^{^{\prime }}$ must be negligibly small along the node direction. Thus,
the interlayer pairing must be strongly momentum dependent in the CuO$_2$
plane. Here, a closely related problem is the magnitude of the interlayer
pairing. In our calculation, we have assumed $\frac{\left| \Delta
_k^{^{\prime }}\right| }{\Delta _0}=1$. This may seem arbitrary at first
sight. However, if we assume that the superconducting pairing originates
from the exchange of the antiferromagnetic spin fluctuation\cite{25,26},
especially, the $\left( \pi ,\pi \right) $ resonance\cite{27}, then both the
magnitude and the momentum dependence of the interlayer pairing can be
easily understood. Since the resonance occurs only in the odd channel, the
spin fluctuation mediating the intralayer and the interlayer pairing have
the same propagator except for an overall sign( note that $\chi
^{(even)}=\chi ^{(intralayer)}+\chi ^{(interlayer)}$, $\chi ^{(odd)}=\chi
^{(intralayer)}-\chi ^{(interlayer)}$). Hence it is quite reasonable that
the intralayer and the interlayer pairing have comparable magnitudes (but
different symmetry). At the same time, since the $\left( \pi ,\pi \right) $
resonance is sharply peaked at $\left( \pi ,\pi \right) $\cite{4,5}, only
momentum close to the hot spots is significantly affected by the interlayer
pairing. Hence the interlayer pairing must be strongly momentum dependent.

Interestingly, this pairing mechanism also naturally explains the different
symmetry of the intralayer and the interlayer pairing. This difference comes
from the overall sign change between the intralayer and the interlayer spin
fluctuation propagator. Since the exchange of intralayer antiferromagnetic
spin fluctuation favors d-wave intralayer pairing\cite{25,26}, or, $\Delta
_{k+\left( \pi ,\pi \right) }=-\Delta _k$, the interlayer pairing mediated
by the interlayer spin fluctuation must have s-wave symmetry, or, $\Delta
_{k+\left( \pi ,\pi \right) }^{^{\prime }}=\Delta _k^{^{\prime }}$. When $%
\Delta _k^{^{\prime }}$ is transformed into the real space, we will see the
interlayer pairing exists only between sites of the same magnetic sublattice
on the two CuO$_2$ plane.s According to our discussion concerning the
relation between the momentum dependence of the magnetic response and the
internal structure of the Cooper pair, such interlayer pairing will enhance
the odd channel response and suppress the even channel response, as we have
observed.

In conclusion, we find the sinusoidal modulation of the $\left( \pi ,\pi
\right) $ resonance observed in experiments can be explained with the
inclusion of the interlayer pairing in the theory. We find the interlayer
pairing has s-wave symmetry in the CuO$_2$ plane and has comparable
magnitude with the d-wave intralayer pairing. We also find the interlayer
pairing has strong momentum dependence and mainly affects momentum close to
the hot spots on the Fermi surface. We find these characteristics of the
interlayer pairing can be understood in a model in which the superconducting
pairing comes from the exchange of the antiferromagnetic spin fluctuation,
especially, the $\left( \pi ,\pi \right) $ resonance.

The author would like to thank T.K.Lee, X.G.Wen, S.C.Zhang and F.C.Zhang for
their valuable comments and Z. Tesanovic for pointing out Ref. 19 to him..

\begin{center}
\newpage FIGURES
\end{center}

FIG. 1. The Fermi surface and the important momentum in our discussion. VHS
denotes the Van Hove singularity.

\medskip

FIG. 2. The imaginary(a) and the real part(b) of the bare susceptibility,
for $\frac{\left| \Delta _k^{^{\prime }}\right| }{\Delta _0}=1$, $\sqrt{%
\Delta _0^2+\left| \Delta _k^{^{\prime }}\right| ^2}=35$ meV. The Fermi
surface is 34 meV above the VHS. Note both the maximal gap and the chemical
potential are the same as that used in Ref. [18]. The intersections of the
straight lines and the curves in (b) give the resonance energies in the odd
and the even channel, for J=150 meV, J$_p$=0.1 J.

\medskip

FIG. 3. The susceptibility after the RPA correction, for J=150 meV, J$_p$%
=0.1 J.

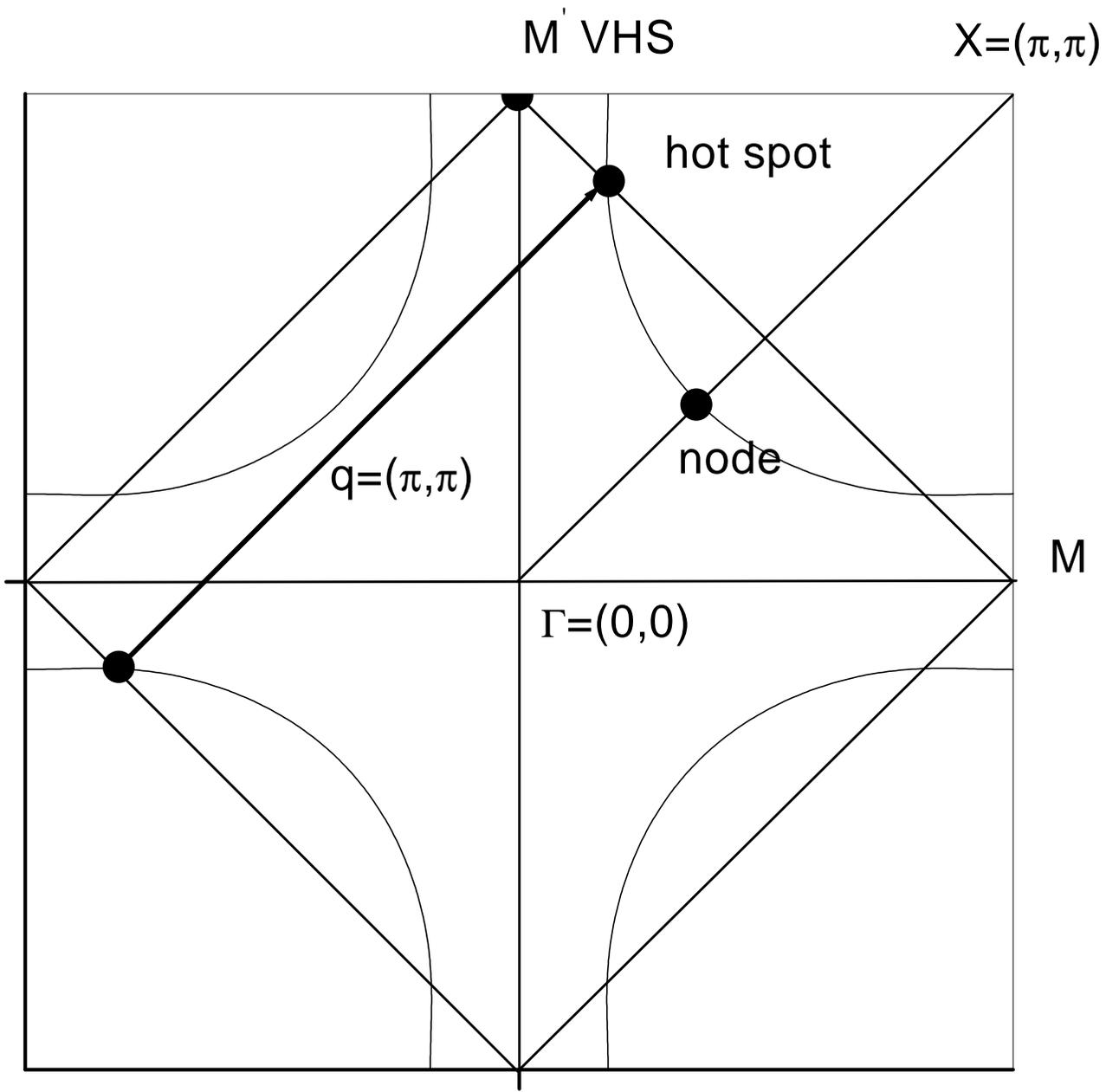
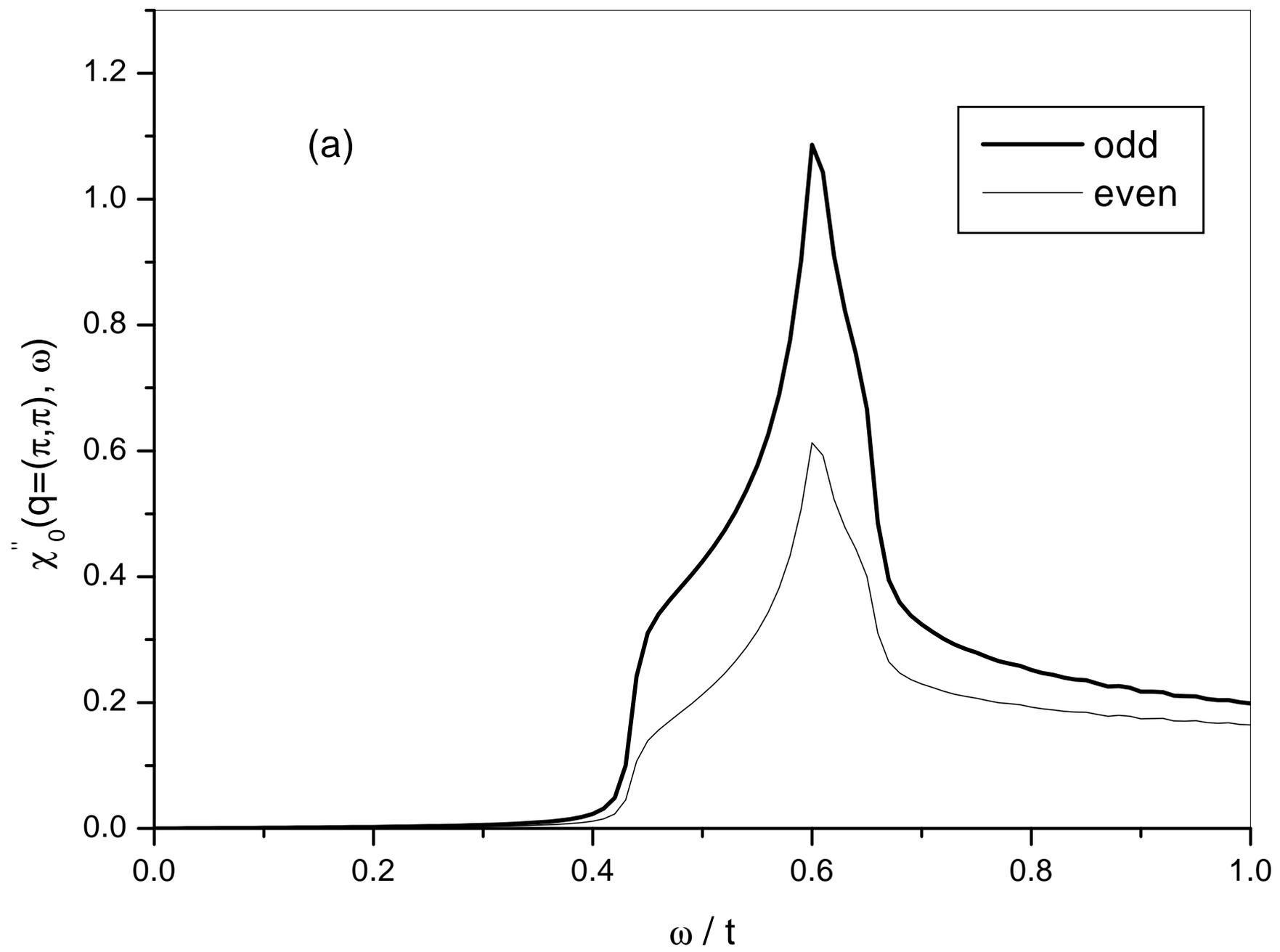
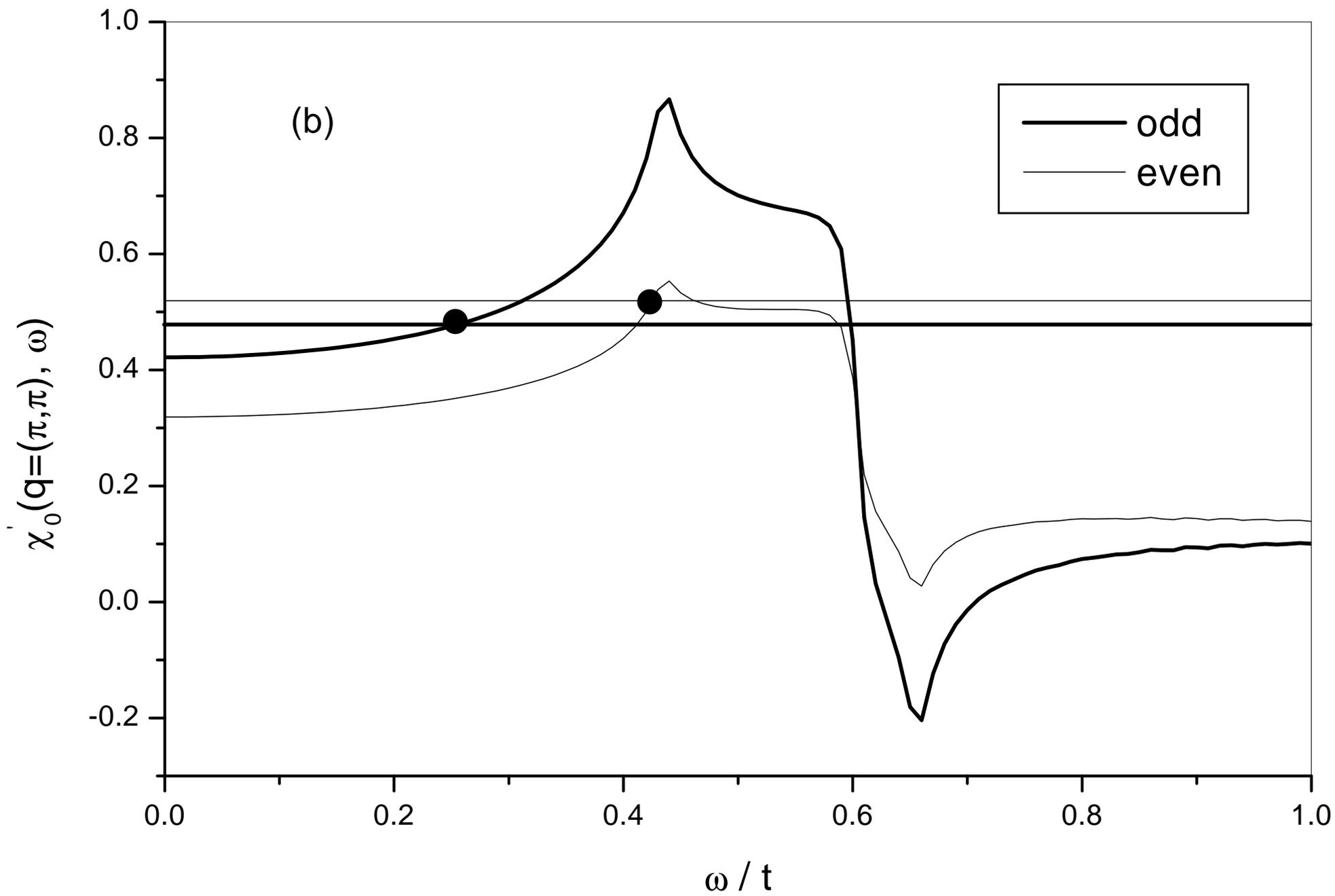
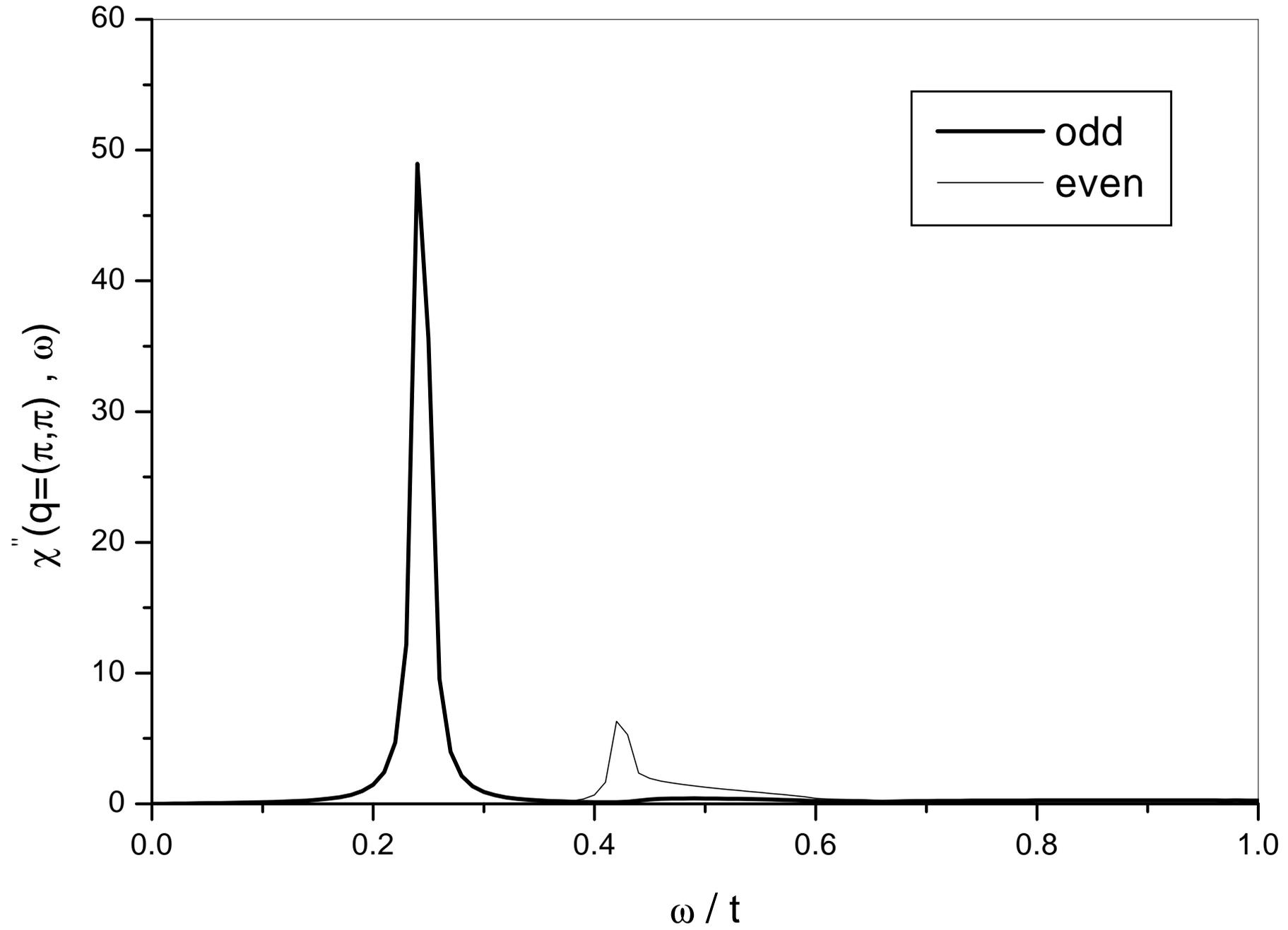
\end{document}